\documentclass[12pt]{iopart}
\usepackage{graphicx}

\begin{document}

\title[Modulation stabilization of Bloch oscillations...]{Modulation stabilization of Bloch oscillations of
two-component Bose-Einstein condensates in optical lattices}
\author{Huai-Qiang Gu$^{1}$ \footnote{ Email: guhq06@lzu.edu.cn},
Jun-Hong An$^{2}$ and Kang Jin$^{3}$}
\address{$^{1}$School of Nuclear Science and Technology, Lanzhou
University, Lanzhou 730000, China}
\address{$^{2}$Center for Interdisciplinary Studies, Lanzhou
University, Lanzhou 730000, China}
\address{$^{3}$Department of Physics, Northwest University, Xi'an
710118, China}

\begin{abstract}
We study the Bloch oscillations (BOs) of two-component Bose-Einstein
condensates (BECs) trapped in spin-dependent optical lattices. Based
on the derived equations of motion of the wave packet in the basis
of localized wave functions of the lattice sites, the damping effect
induced by the intercomponent and intracomponent interactions to the
BOs is explored analytically and numerically. We also show that such
damping of the BOs can be suppressed entirely if all the atom-atom
interactions are modulated synchronously and harmonically in time
with suitable frequency via the Feshbach resonance. When the
intercomponent and the intracomponent interactions have inverse
signs, we find that the long-living BOs and even the revival of the
BOs can be achieved via only statically modulating the configuration
of optical lattices. The results provide a valuable guidance for
achieving long-living BOs in the two-component BEC system by the
Feshbach resonances and manipulating the configuration of the
optical lattices.
\end{abstract}


\maketitle

\section{Introduction}
The system of ultracold atomic gases in optical lattices has become
a nice experimental platform to simulate the effects in condensate
matter physics \cite{Morsch2006}. The well controllability of such
system makes many sophisticated effects in condensed matter physics
be well studied in this system \cite{Bloch2005}. The Bloch
oscillation (BO) is the oscillatory motion of a quantum particle in
a periodic potential when it is subjected to an external force. It
was originally predicted in solid-state system where the motions of
the electrons in tilted periodic potentials undergo coherent
oscillations \cite{Bloch1929}. The formal resemblance between
electrons in crystals and Bose-Einstein condensates (BECs) in
optical lattices has inspired an extensive interest to explore the
BOs in optical lattice system. The successful observations of BOs
have been reported for atoms in interacting BECs
\cite{BenDahan,Anderson,Morsch,Salger} and for ensembles of
noninteracting quantum-degenerate fermions \cite{Roati} in tilted
optical lattices.

However, the perfect BOs can only be available in the ideal case
where there is no interactions among the atoms of BECs. Practically,
due to the intrinsically weak interactions of atoms, the momentum
distribution of the BECs will show a rapid broadening, which causes
the atoms of BECs to lose their phase coherence, i.e. the dephasing
effect \cite{Trombettoni,Gaul}. Consequently, the BOs in BECs cannot
persist on for a long time. In the framework of the mean-field
treatment, the motion of the BECs can be well described by the
so-called Gross-Pitaevskii equation (GPE) with a nonlinear term. It
is believed that such nonlinearity induced by the atom-atom
interactions generally leads to a breakdown of the BOs, as recently
studied experimentally \cite{Gustavsson2008} and theoretically
\cite{Gangardt,Krimer2009,Kolovsky2010}. Therefore, a natural
question is: Is it possible to prolong or even stabilize the BOs by
some active control ways?

Addressing this question, some progress, especially in
single-component BECs, has been made. It has been found that a
long-living BO can be induced by properly designing the spatial
dependence of the scattering length \cite{Salerno} and the
configuration of the optical lattices \cite{Walter10}. Gustavsson
{\it et al.} \cite {Gustavsson2008} showed the experimental evidence
that the dephasing time of the BOs can be much enhanced by decreasing
the interaction strengths via the Feshbach resonance \cite{Donley}.
However, in many situations, such finite enhancement to the dephasing
time of BOs is not enough, and one is desired to preserve the BOs
forever. Recently, Gaul {\it et al.} have reported that a persistent
BO of single-component BECs can be obtained by modulating
interaction harmonically in time with suitable frequency and phase
\cite{Gaul,Diaz10}, which can be easily realized by means of
Feshbach resonance.

So far, most of the studies of the stabilization control to BO of
BECs in optical lattice are based on the single-component BEC case.
Compared with single-component BEC, the two-component BEC system may
exhibit more novel physical effects due to the condensate mixtures
\cite{Pu,Ostrovskaya,Xiuquan Ma,Jian-Jun Wang}. In this system, both
intracomponent and intercomponent atom-atom interactions take
effects upon the nonlinear behavior of the BECs. The two-component
BECs can be experimentally realized by spin-dependent optical
lattices, which hold the BECs with the two components composed of
two distinct hyperfine states of the same atomic species
\cite{Mandel}.

In the present paper, we study systematically the modulation
stabilization of the BOs of two-component BECs in optical lattices.
We mainly use two control ways, one is by modulating periodically
the interactions via Feshbach resonance; the other is by tuning
statically the parameters of the optical lattices. We will show that
the stable BOs can be obtained when the interactions are modulated
synchronously and harmonically in time with suitable frequencies.
Moreover, if the intercomponent and the intracomponent interactions
have inverse signs, the long-living BOs and even the revival of the
BOs can be achieved via only tuning the the relative separation
between lattices.

The paper is organized as follows. In Sec. II, we discuss the
methods and formalism used in this work. In Sec. III, we explore
quantum manipulation of BOs from two aspects according the signs of
the intracomponent atom-atom interactions. Finally, a summary is given in Sec. IV.

\section{Model and formulation}
\subsection{Gross-Pitaevskii equations for the two-component BECs in an optical lattice }
We consider two-component BECs which are composed of bosonic atoms
of the same isotope but having different internal spin states, e.g.
$^{87}$Rb atoms in hyperfine states $|F=2,m_F = 2\rangle$ and $|F =
1,m_F = -1\rangle$ \cite{Schmaljohann}. The BECs are trapped in
spin-dependent optical lattices. The dynamics of the system is
governed by the coupled GPEs under the mean field approximation,
\begin{equation}
i\hbar\partial_t\Phi_{i}=\big[-\frac{\hbar^{2}}{2M}\nabla^{2}+V_{i}+\sum_{j=1}^{2}{g_{ij}(t)|\Phi_{j}|^{2}}\big]\Phi_{i},
\label{gpe}
\end{equation}
where $\Phi_{i}$ ($i=1,2$) is the macroscopic condensate order
parameter of the $i$-th component with identical mass $M$. The
time-dependent interaction coefficients are given by
$g_{ij}(t)=4\pi\hbar^2a_{ij}(t)/M$ with $a_{ij}(t)$ being the
$s$-wave scattering length which can be controlled via the Feshbach
resonance induced by the modulated magnetic field . The external
potential felt by the $i$-th component can be decomposed into
$V_{i}=V_c+V_{Li}$, where $V_{c}=fz$ is a linear potential induced
by a constant force $f$ and $V_{Li}$ is trapping potential of the
optical lattice. The additional weak potential $V_{c}$ tilts the
optical potentials and drives coherent oscillations \cite{Anderson}.
The trapping potentials for different components can be explicitly
written as
 $V_{L1}=U_{p}\sin^2(k_{L}z+\frac{\theta}{2})$ and
$V_{L2}=U_{p}\sin^2(k_{L}z-\frac{\theta}{2})$, where $U_{p}$ is the
depth of the 1D lattice potentials, $k_{L}$ is the wave vector of
the lasers used to construct the optical lattice, and $\theta$ is
the polarization angle of the two counterpropagating laser beams to
form the standing wave configuration of the optical lattice \cite
{Ostrovskaya,Mandel}. By changing $\theta$, one can also control the
separation between the two potentials.

When the linear field is too weak to induce Landau-Zener tunneling
\cite{Cristiani,Witthaut}, BO can be described by an adiabatic
evolutions of the BECs in the lowest lattice band. In collective
coordinates \cite{Trombettoni}, the condensate order parameter
$\Phi_{i}(r,t)$ can be expanded as a linear combination of the wave
packets localized at the individual lattice sites, i.e. the Wannier
wave functions $\phi_{n_i}(r)$, as
\begin{equation}
\Phi_{i}(r,t)=\sqrt{N_{i}}\sum_{n_{i}}{\psi_{i,n_{i}}(t)\phi_{n_{i}}(r)}\label{wne},
\end{equation}
where $N_{i}$ is the total number of particles of the $i$-th
component and the Wannier wave function $\phi_{n_{i}}$ satisfies the
orthogonality condition $\int{dr\phi_{n_{i}}\phi_{n_{i}\pm1}}=0$,
and the normalization condition $\int{dr\phi_{n_{i}}^2}=1$.
$\psi_{i,n_{i}}=\sqrt{N_{i,n_{i}}(t)/N_{i}}e^{i\theta_{i,n_{i}}(t)}$,
where $N_{i,n_{i}}(t)$ and $\theta_{i,n_{i}}(t)$ are the number of
particles and phase, respectively, is the amplitude of the $i$-th
component trapped in the site $n_{i}$. In the following, we assume
that the two components have the same total number of particles,
i.e., $N_{1}=N_{2}$. Substituting Eq. (\ref{wne}) into Eqs.
(\ref{gpe}), we can discretize Eqs. (\ref{gpe}) into a set of
coupled nonlinear equations with respect to different lattice site
$n_i$,
\begin{eqnarray}
&&i\dot\psi_{i,n_{i}}=-\frac{\psi_{i,n_{i}-1}+\psi_{i,n_{i}+1}}{2}+\big[\epsilon_{i,n_{i}}+\Lambda_{ii}(t)|\psi_{i,n_{i}}|^2
\nonumber\\
&&~~+\Lambda_{ij}(t)(\eta_\tau|\psi_{j,n_{i}+\tau}|^2+\eta_{\tau-1}|\psi_{j,n_{i}+\tau-1}|^2)\big]\psi_{i,n_{i}}\label{eq4},
\end{eqnarray}
where the overdot denotes the time derivative and $i\neq{j}$ labels
the two different components of BECs.
$\epsilon_{i,n_{i}}=\frac{1}{2J}\int{dr[\frac{(\hbar\nabla\phi_{n_{i}})^2}{2M}+V_{i}\phi_{n_{i}}^2]}$
with $J=-\int
dr[\frac{\hbar^2}{2M}\nabla\phi_{n_{i}}\nabla\phi_{n_{i}+1}+
\phi_{n_{i}}V_{i}\phi_{n_{i}+1}]$ being the tunnel parameter.
$\Lambda_{ii}=\frac{g_{ii}(t)N_{i}}{2J}\int{dr\phi_{n_{i}}^4}$
describes the intracomponent atom-atom interaction strength.
$\Lambda_{ij}=\frac{g_{ij}(t)N_{j}}{2J}$ multiplying with
$\eta_\tau=\int{dr\phi_{n_{i}}^2\phi_{n_{i}+\tau}^2}$ and
$\eta_{\tau-1}=\int{dr\phi_{n_{i}}^2\phi_{n_{i}+\tau-1}^2}$ describe
the intercomponent atom-atom interaction strengths, where
$\eta_{\tau}$ and $\eta_{\tau-1}$ stem from the overlaps of the wave
functions of the two components. $\tau$ (in the lattice unit) is
determined by the relative separation between the two nearest
neighboring spin-dependent potentials, which can be controlled by
the polarization angle $\theta$. It is noted that not only all the
interactions are time-dependent but also intercomponent nonlinear
interactions depend on $\eta_{\tau}$ and $\eta_{\tau-1}$. In Eq.
(\ref{eq4}) the time has been rescaled to be dimensionless as
$t\rightarrow\hbar t/2J$.

In 1D optical potentials, we can denote the Wannier wave function as
$\phi_{n_{i}}(r)=\phi(x,y)\phi_{n_{i}}(z)$. The transverse wave
function can be expressed as a 2D Gaussian profile
$\phi(x,y)=\phi_0(x)\phi_0(y)$, where
$\phi_{0}(\alpha)=\frac{1}{\sqrt[4]{\pi}\sqrt{\sigma_{\alpha}}}\exp(-\frac{\alpha^2}{2\sigma_{\alpha}^2})$
with $\sigma_{\alpha}$ being the Gaussian widths in the $\alpha=x,y$
directions. The wave function along the direction of the optical
lattice can be denoted as $\phi_{n_{i}}(z)=\phi_{0}(z-n_{i}d)$ with
$d=\pi/k_{L}$ being the lattice constant. Based on the variational
ansatz for $\phi_{n_{i}}(z)$, a minimum energy can be obtained when
the width of $\phi_{n_{i}}(z)$ equals to
$\sigma_{z}=\frac{d}{\pi}\sqrt[4]{U_{p}/E_{rec}}$, where $E_{rec}$
is the recoil energy \cite{Cristiani}. Under this consideration, the
parameters in Eq. (\ref{eq4}) can be determined explicitly as
\begin{eqnarray}
&\epsilon_{1,n_{1}}&=\omega n_1+\frac{\theta^2{U_{p}}}{8},~~\epsilon_{2,n_{2}}=\omega n_2-\frac{\theta^2{U_{p}}}{8},\nonumber\\
&\Lambda_{ii}(t)&=\frac{1}{2J}\frac{g_{ii}(t)N_{i}}{(2\pi)^{3/2}\sigma_{x}\sigma_{y}\sigma_{z}},~~\eta_{\tau}=\exp(-\frac{\tau^2d^2}{2\sigma_{z}^2}),\nonumber\\
&J=&\exp(\frac{d^2}{4\sigma_{z}^2})\{\frac{\hbar^2}{2M}\frac{d^2-2\sigma_{z}^2}{4\sigma_{z}^4}-\frac{U_{p}}{2}[1+\exp(\frac{-\pi^2\sigma_{z}^2}{d^2})]\},\label{prm}
\end{eqnarray}
where
 $\omega=\frac{fd}{2J}$ and $f$ corresponds to the weak atomic gravity. From Eq. (\ref{eq4}) and the canonical equation
$\dot{\psi_{i}}=\frac{\partial\mathcal{H}}{\partial{(i\psi_{i}^*)}}$,
the Hamiltonian functions can be obtained
\begin{eqnarray}
\mathcal{H}_{i}&=&\sum_{n_i}\big\{-\frac{\psi_{i,n_{i}}\psi_{i,n_{i}+1}^*+\psi_{i,n_{i}}^*\psi_{i,n_{i}+1}}{2}+\big[\epsilon_{i,n_{i}}
\nonumber\\
&&+\frac{\Lambda_{ii}(t)}{2}|\psi_{i,n_{i}}|^2+\Lambda_{ij}(t)(\eta_{\tau}|\psi_{j,n_{i}+\tau}|^2
\nonumber\\
&&+\eta_{\tau-1}|\psi_{j,n_{i}+\tau-1}|^2)\big]|\psi_{i,n_{i}}|^2\big\},\label{hmt}
\end{eqnarray}
both the Hamiltonian $\mathcal{H}_{i}$ and the norm
$\sum_{n_{i}}|\psi_{i,n_{i}}|^{2}=N_{i}$ are conserved.

\subsection{The dynamics of wave packet: Bloch oscillations}
In order to analyze how the interactions affect the BOs of the
two-component BECs, we parameterize the Gaussian profile wave packet
for $i$-th component as \cite{Trombettoni}
\begin{equation}
\psi_{i,n_{i}}=\sqrt{K_{i}}\exp[-\frac{(n_{i}-\xi_{i})^2}{\gamma_{i}^2}+ip_{i}(n_{i}-\xi_{i})+i\frac{\delta_{i}}{2}(n_{i}-\xi_{i})^2].
\label{wpe}
\end{equation}
where $K_{i}=\sqrt{\frac{2}{\pi\gamma_{i}^2}}$ is a normalization
factor. The Gaussian wave packet is parameterized by four types of
quantities: the center-of-mass position $\xi_i(t)$, the width of the
wave packet described by $\gamma_i(t)$, the linear phase $p_i(t)$
describing the group velocity of the wave packet, and the quadratic
phase $\delta_i(t)$ over the wave packet. The latter phase allows
us, on the one hand, to describe the linear evolution of the wave
packet for which the quadratic dispersion in momentum space directly
translates into a quadratic phase in real space. On the other hand,
the nonlinearity due to the atom-atom interactions also leads to a
quadratic phase since the density near the Gaussian maximum is
quadratic \cite{Morsch2006}. Such Gaussian profile wave packet was
used to explain successfully the BO in Anderson-Kasevich experiment
\cite{Anderson}.

The dynamical evolution of the wave packet can be obtained by a
variational principle from the Lagrangian
$\mathcal{L}_{i}=i\sum_{n_{i}}\dot{\psi}_{i,n_{i}}\psi_{i,n_{i}}^*-\mathcal{H}_{i}$.
After some algebra, the Lagrangian can be achieved
\begin{eqnarray}
 \mathcal{L}_{i}&=&p_{i}\dot{\xi_{i}}-\frac{\gamma_{i}^2\dot{\delta_{i}}}{8}
+e^{-\chi_{i}}\cos{p_{i}}-\frac{\Lambda_{ii}(t)}{2\sqrt{\pi}\gamma_{i}}-v_{i}
\nonumber\\
&&-\Lambda_{ij}(t)\frac{\kappa}{\sqrt{\pi}}[\eta_{\tau}e^{-\mu_{\tau}}
+\eta_{\tau-1}e^{-\mu_{\tau-1}}],\label{eq7}
\end{eqnarray}
where
$\chi_{i}=\frac{1}{2\gamma_{i}^2}+\frac{\gamma_{i}^2\delta_{i}^2}{8}$,
$v_{i}=K_{i}\int
dn_{i}\epsilon_{i,n_{i}}\exp[\frac{-2(n_{i}-\xi_{i})^2}{\gamma_{i}^2}]$,
$\kappa=\frac{\sqrt{2}}{\gamma}$ with
$\gamma^2=\gamma_{1}^2+\gamma_{2}^2$, and
$\mu_{\tau}=\kappa^2\xi_{\tau}^2$ with
$\xi_{\tau}=\xi_{i}-\xi_{j}+\tau$. It is noted that in our
calculation the summation over $n_i$ has been replaced by
integration when the widths $\gamma_i$ are not too small
\cite{Trombettoni,Jian-Jun Wang}. The equations of motion of the
collective coordinates can be obtained from the Euler-Lagrange
equations
\begin{eqnarray}
\dot{\xi_{i}}&=&e^{-\chi_{i}}\sin{p_{i}},\label{xi}\\
\dot{\gamma_{i}}&=&\gamma_{i}\delta_{i}e^{-\chi_{i}}\cos{p_{i}},\label{gm}\\
\dot{p_{i}}&=&\frac{2\kappa^3\Lambda_{ij}(t)}{\sqrt{\pi}}
[\eta_{\tau}\xi_{\tau}e^{-\mu_{\tau}}+\eta_{\tau-1}\xi_{\tau-1}e^{-\mu_{\tau-1}}]-\omega,
\label{p}\\
\dot{\delta_{i}}&=&(\frac{4}{\gamma_i^4}-\delta_i^2)e^{-\chi_{i}}\cos{p_{i}}+\frac{2\Lambda_{ii}(t)}{\sqrt{\pi}\gamma_{i}^3}
+\frac{\kappa^5z\Lambda_{ij}(t)}{\sqrt{\pi}}, \label{dt}
\end{eqnarray}
where
$z=\eta_{\tau}(\gamma^2-4\xi_\tau^2)e^{-\mu_{\tau}}+\eta_{\tau-1}(\gamma^2-4\xi_{\tau-1}^2)e^{-\mu_{\tau-1}}$.

To highlight the essential physics, from the coupled Eqs.
(\ref{xi})-(\ref{dt}), the equation of motion of the center of the
wave packet can be recast into
\begin{equation}
\ddot{\xi_{i}}+\alpha_{i}\dot{\xi_{i}}=\beta_{i}, \label{xieom1}
\end{equation}
where
\begin{eqnarray}
\alpha_{i}&=&\frac{\delta_{i}\Lambda_{ii}(t)}{2\sqrt{\pi}\gamma_{i}}+\frac{\kappa^5\gamma_{i}^2\delta_{i}\Lambda_{ij}(t)z}{4\sqrt{\pi}},\label{alpha}\\
\beta_{i}&=&\dot{p_{i}}e^{-\chi_{i}}\cos{p_{i}}. \label{beta}
\end{eqnarray}
It is noted that Eq. (\ref{xieom1}) can recover the equation of
motion of the wave-packet center for single component BEC under the
condition $\Lambda_{ij}=0$ as \cite{Trombettoni}
\begin{equation}
\ddot{\xi_{i}}+\alpha_{i}\dot{\xi_{i}}+\omega^2\xi_{i}=\frac{\omega
\Lambda_{ii}}{2\sqrt{\pi}}[\gamma_{i}^{-1}(0)-\gamma_{i}^{-1}(t)],
\label{xieom2}
\end{equation}
which is a standard equation of motion for a harmonic oscillation
with an effective damping rate $\alpha(t)$. Under the ideal
condition $\Lambda_{ii}=0$, $\alpha(t)=0$ and the system undergoes a
perfect oscillation, which is the perfect BOs with the frequency
$\omega$ as a result of the driven field. The damping of the BOs is
caused by the intracomponent interactions $\Lambda_{ii}$, which
means that the nonlinearities inevitably lead to the breakdowns of
the BOs. In the situation of two-component BECs, the intercomponent
interaction $\Lambda_{ij}$ will add another nonlinearity to each
individual component of the BECs, which gives an additional
contribution to the damping rate, as shown in Eq. (\ref{alpha}).
Besides the damping rate, the intercomponent interaction also exerts
an effective driven force to BOs. Explicitly, under the condition
that $\Lambda_{ij}$ is small compared with the constant force $f$
felt by the BECs, one can rewritten Eq. (\ref{xieom1}) as
\begin{equation}
\ddot{\xi_{i}}+\alpha_{i}\dot{\xi_{i}}+\widetilde{\omega}^2\xi_{i}=\eta_{i}(\Lambda_{ii},\Lambda_{ij}),
\label{xieom3}
\end{equation}
which shows that the perfect BOs of the system are distorted by the
cooperative action of the effective damping rate $\alpha(t)$ and the
effective driven force $\eta_{i}(\Lambda_{ii},\Lambda_{ij})$. It is
noted that the frequency $\widetilde{\omega}$ of the BOs, which
corresponds to the inverse of the right-hand side of Eq. (\ref{p}),
for the two-component case is slightly detuned from $\omega$ by the
intercomponent interactions $\Lambda_{ij}$. The explicit analysis of
the BOs governed by $\alpha(t)$ and
$\eta_{i}(\Lambda_{ii},\Lambda_{ij})$ will be shown by the
quantitative calculations with the experimentally adjustable
parameters in the next section.

\section{Quantum manipulation of Bloch oscillation in the optical lattice}
From the above analytical results, we can see that the
nonlinearities contributed from both of the intercomponent and
intracomponent interactions generally lead to the breakdowns of the
BOs of the BECs in the optical lattices. Therefore, the interaction
coefficients $g_{ij}$ of such nonlinearities, which are essentially
determined by the $s$-wave scattering lengths $a_{ij}$, have
profound impact on the dynamics of two-component BECs in the optical
lattices. The magnitude and sign of these parameters sensitively
influence dynamical behaviors of this ultracold boson system. In
cold-atom experiments, Feshbach resonance is a quite effective
mechanism that can be used to modulate $g_{ij}$. Inspired by a
recent experimental investigation of the ultracold molecule
production via a sinusoidal magnetic field modulation to the
interaction coefficient around the Feshbach resonance
\cite{Thompson05}, we intend to explore the possibility to stabilize
the BOs via such periodic modulations to the interaction
coefficients $g_{ij}$ in the following. Besides the magnetic field
induced Feshbach resonance, another way to modulate the dynamics of
the BECs in our system is via adjusting the configuration of the
optical lattices. The separation between the spin-dependent
potentials felt by the two components of BECs, which can be adjusted
by the polarization angle $\theta$ of the lasers, essentially
determines the intercomponent interactions of the BECs. We also
examine the influence of the separation on the dynamics of BOs. The
combined effect of the Feshbach resonance and a periodic external
potential has been widely studied
\cite{Abdullaev,Baizakov,Brazhnyi}.

\subsection{$a_{ii}(0)>0,a_{ij}(0)>0$}
In this case, both of the intercomponent and intracomponent
interactions are repulsive. Without loss of generality, we assume
that the intercomponent interactions $\Lambda_{ij}$ relate to the
intracomponent interactions $\Lambda_{ii}$ as:
$\Lambda_{12}=\Lambda_{21}=\sqrt{\Lambda_{11}\Lambda_{22}}$ in our
numerical simulations.
\begin{figure*}[tbp]
\centering
\includegraphics[width = \columnwidth]{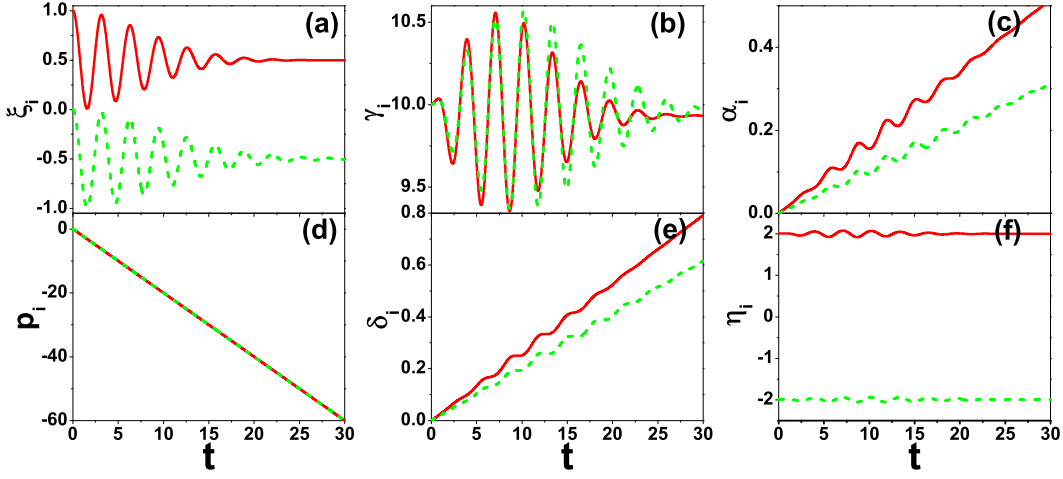}
\caption{(Color online) Attenuations of BOs without modulating the
interactions. Solid and dashed lines indicate two individual
components with the intracomponent interactions as $\Lambda_{11}=20$
and $\Lambda_{22}=15$, respectively. The other parameters are chosen
as: $U_{p}=16E_{rec}$, $\omega=2$, and $\tau=0.5$. The initial
conditions are set as: $p_{1}(0)=p_{2}(0)=0$,
$\delta_{1}(0)=\delta_{2}(0)=0$, $\xi_{1}(0)=1$, $\xi_{2}(0)=0$,
$\gamma_{1}(0)=\gamma_{2}(0)=10$.} \label{fig:1}
\end{figure*}

\begin{figure*}[tbp]
\centering
\includegraphics[width = \columnwidth]{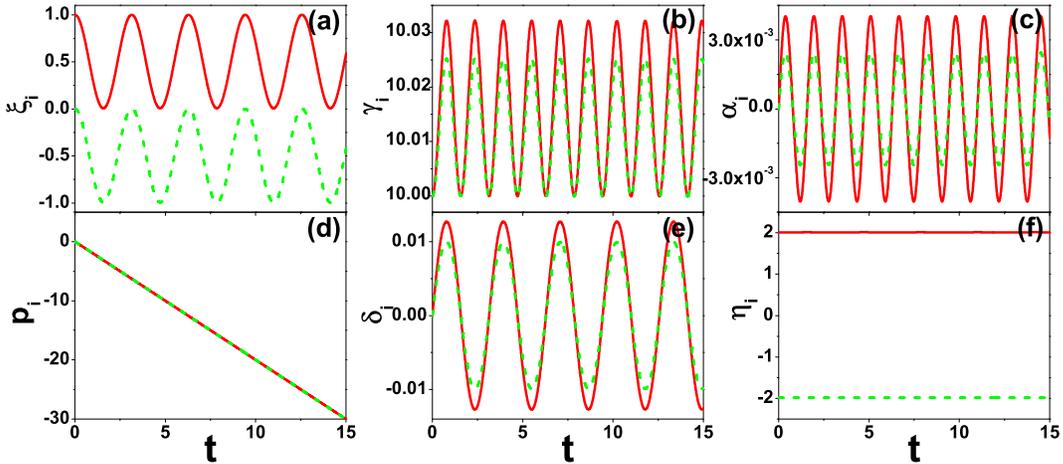}
\caption{(Color online) The stabilization of BOs by modulating the
interactions. All interactions are modulated by $\cos(\omega t)$.
The same parameters and notation are used as Fig. \ref{fig:1}.}
\label{fig:2}
\end{figure*}

To see the effect of nonlinearities on the BOs, we plot in Fig.
\ref{fig:1} the attenuations of BOs without modulations. From the
time-dependent behaviors of the wave-packet centers $\xi_{i}$ [Fig.
\ref{fig:1}(a)] we can see clearly the breakdown of BOs with time
evolution. Such damping oscillations are manifested by the behaviors
of the effective damping rates $\alpha_i$ [Fig. \ref{fig:1}(c)] and
the driven force $\eta_i$ [Fig. \ref{fig:1}(f)], where $\alpha_i$
are always positive and increase with time, and $\eta_{i}$ tend to
constants after the same cycles as $\xi_{i}$. Compared with the case
for the single-component BECs \cite{Trombettoni}, the presence of
the intercomponent interactions, sharing the same sign with
intracomponent ones here, play the role as an additional
nonlinearities and speed up the collapses in our two-component
situation. So, the stronger the intercomponent interactions are, the
faster the BOs damp. Fig. \ref{fig:1}(d) shows that $p_{i}$, the
associated momenta of $\xi_{i}$, increase linearly with time. This
can be understood from the analysis of Eq. (\ref{p}). The first term
of the right hand of Eq. (\ref{p}), which is contributed from the
intercomponent interactions, is much smaller than the second term,
which is contributed from the linear potential. Consequently, the
time-dependent behaviors of $p_i$ are dominated by the second term,
i.e. $p_i(t)\approx p_i(0)-\omega t$. Fig. \ref{fig:1}(b) shows that
the widths $\gamma_{i}$ undergo breathing oscillations and soon
approach constants. The associated momenta $\delta_i$ are also
divergent, as shown in Fig. \ref{fig:1}(e). All these time-dependent
behaviors indicate that the system is set into a macroscopical
quantum self-trapping mode \cite{Smerzi,Milburn,Tristram} due to the
nonlinearities from the intercomponent and intracomponent
interactions.

Now we focus on how to stabilize the BOs in our system. We mainly
use the way by modulating periodically the interactions via a
magnetic field.

In Fig. \ref{fig:2} we plot the time-dependent behaviors of the
wave-packet variables under a $\cos \omega t$, where $\omega$ is the
frequency of BOs, modulation to all the interactions. We can see
that the damping of the BOs can be fully stabilized by such
modulation and perfect oscillations is thus obtained. Such
persistent phenomena can be also explained straightforwardly by
studying the time-dependence of the damping coefficients
$\alpha_{i}$ [Fig. \ref{fig:2}(c)] and the effective driven forces
$\eta_{i}$ [Fig. \ref{fig:2}(f)]. In contrast to the positivity in
the full evolution range in Fig. \ref{fig:1}(c), $\alpha_{i}$ in
this case exhibit periodic oscillations between positive and
negative values with definite amplitudes, which characterizes well
the lossless BOs of $\xi_i$. It means that the effective damping
coefficients with alternate signs inspirit the system itself to
guarantee the stabilizations of the BOs.  Besides, we find that the
driven forces in Fig. \ref{fig:2}(f) are replaced by constant values
completely after such modulation.  It has been proven that there is
a family of stable solutions in terms of collective coordinates in
single-component BEC system when all the interactions are modulated
by cos($\omega$t) \cite{Gaul}. In fact, in two-component ones, the
coupled terms in the Eqs. (\ref{gpe}) can be regarded as additional
nonlinearities, which possess the same time dependence as
intracomponent interactions so long as modulating all interactions
harmonically in time with the same suitable frequencies. As viewed
from the mean field, the two components can be reduced into two
independent single ones. So it is understandable that such
modulation stabilization also presents in our two-component BECs
system.

\begin{figure*}[tbp]
\centering
\includegraphics[width = \columnwidth]{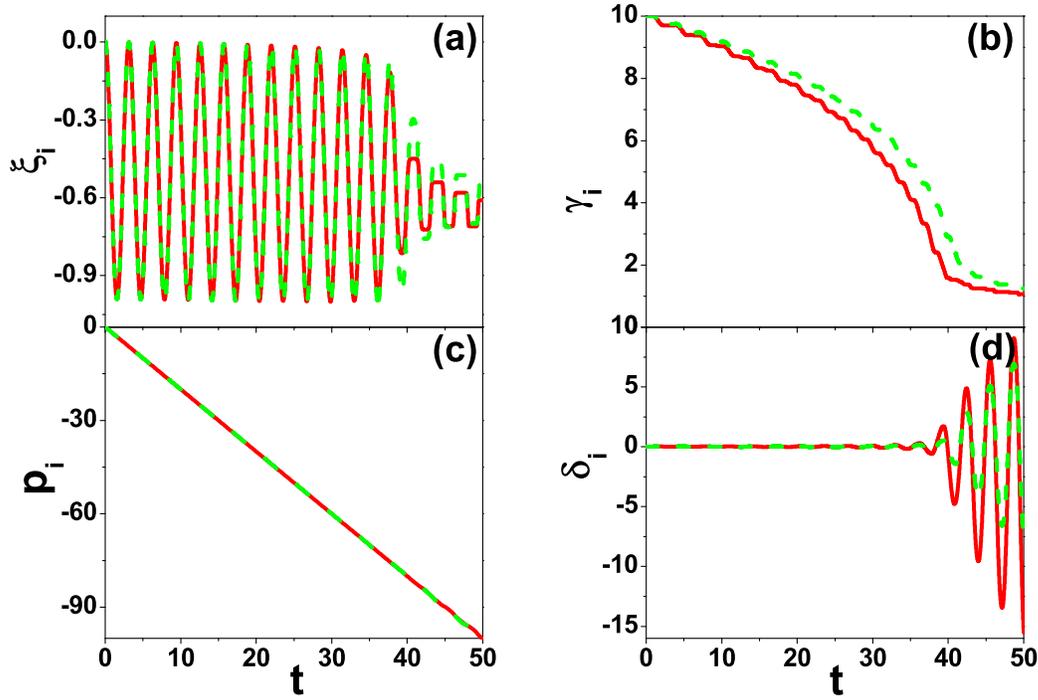}
\caption{(Color online) The attenuations of the BOs with modulating
interactions by $\sin(\omega t)$. $\tau=0.1$ and the same other
parameters and notation as Fig. \ref{fig:1}.} \label{fig:4}
\end{figure*}

It is noted that the effect of the modulation to the BOs is
sensitively dependent of the forms of modulating field we used. In
Fig. \ref{fig:4}, we plot the numerical simulation when all the
interactions are modulated by a $\sin(\omega t)$ field. It is found
that the damping of the BOs manifested by $\xi_i$ is not suppressed
and the BOs is destroyed in several rounds of oscillation. The
wave-packet widths $\gamma_i$ reduce their amplitudes quickly. So
the modulation takes no effect in this case.

In the present case, the intercomponent interactions share the same
sign as the intracomponent ones, so the tuning of the relative
separation $\tau$ does not have constructive action to suppress the
damping of BOs. However, the things are changed dramatically when
the intercomponent interactions have opposite sign to intracomponent
ones, as discussed in the following.

\subsection{$a_{ii}(0)>0,a_{ij}(0)<0$}
In this case, the original intracomponent interactions are
repulsive, while the original intracomponent ones are attractive.
Different to the above case, there are many interesting effects
induced from the competition between such two kinds of atom-atom
interactions. For example, the stability of static solitonic
excitations in two-component BECs have been analyzed within the
Gross-Pitaevskii approximation \cite{Schumayer}. For convenience, we
assume the intercomponent atom-atom interactions to be
$\Lambda_{12}=\Lambda_{21}=-\sqrt{\Lambda_{11}\Lambda_{22}}$ in our
numerical simulations.

\begin{figure*}[tbp]
\centering
\includegraphics[width = \columnwidth]{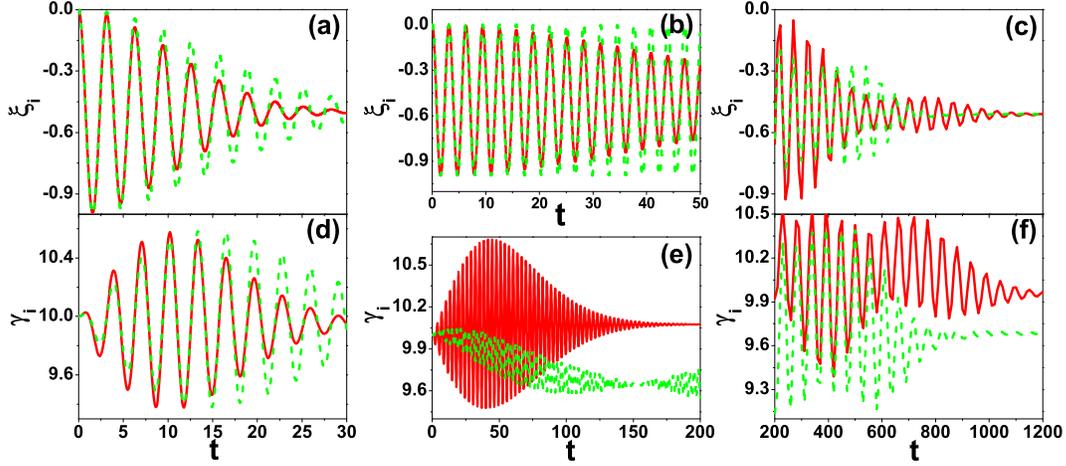}
\caption{(Color online) Attenuations of BOs without modulating
interactions for different potential separations $\tau$. Solid and
dashed lines indicate two individual components with the
intracomponent interactions as $\Lambda_{11}=20$ and
$\Lambda_{22}=18$, respectively. The other parameters are chosen as:
$U_{p}=16E_{rec}$, $\omega=2$, and $\tau=0.5$ for (a, d), $\tau=0.1$
for (b, e), and $\tau=0.05$ for (c, f). The initial conditions are
set as: $p_{1}(0)=p_{2}(0)=0$, $\delta_{1}(0)=\delta_{2}(0)=0$,
$\xi_{1}(0)=\xi_{2}(0)=0$, $\gamma_{1}(0)=\gamma_{2}(0)=10$.}
\label{fig:5}
\end{figure*}

As analyzed in above case, the BOs are destroyed by the
nonlinearities. From this point, the dynamical behaviors of the wave
packet in the present case show no difference to the above one.
However, we can prolong the coherent time of the BOs dramatically by
tuning the relative separation $\tau$ of the potentials felt by the
two components in the present case. To confirm this, we plot the
time evolutions of the wave-packet centers $\xi_{i}$ and widths
$\gamma_{i}$ in Fig. \ref{fig:5} for different relative separations.
A large $\tau$ means a large distances between the nearest neighbors
of the Wannier wave functions, which in turn induces a small
intercomponent interaction rate $\eta_\tau$. Fig. \ref{fig:5}(a,d)
show the breakdown of the BOs when the intercomponent interactions
are small for a large $\tau$. With the decrease of $\tau$, the
intercomponent interactions get stronger. The damping of BOs are
obviously slowed down [Fig. \ref{fig:5}(b,e)]. Especially, it is
noticed that $\gamma_{i}$ shows revival at about $t=150$, as the
dashed line in Fig. \ref{fig:5}(e). It provides a valuable guidance
that the dynamics of the system would show revival in this case. If
the relative separation $\tau$ is further reduced so that the two
lattices are extremely close, the BOs show obvious revival [Fig.
\ref{fig:5}(c,f)]. This phenomenon is caused by the competition
between the intercomponent and intracomponent interactions. Because
the intracomponent interactions have inverse sign with the
intercomponent ones, the nonlinearities contributed from the
intracomponent interactions are partially counteracted by the
intercomponent ones. For the full overlap at
$\tau=0$, the two components perfectly mix together and attenuations
of BOs reappear. To sum up, the coherent time of the BOs can be
much enhanced by only tuning the relative separation $\tau$ of the
optical lattices.

\begin{figure*}[tbp]
\centering
\includegraphics[width = \columnwidth]{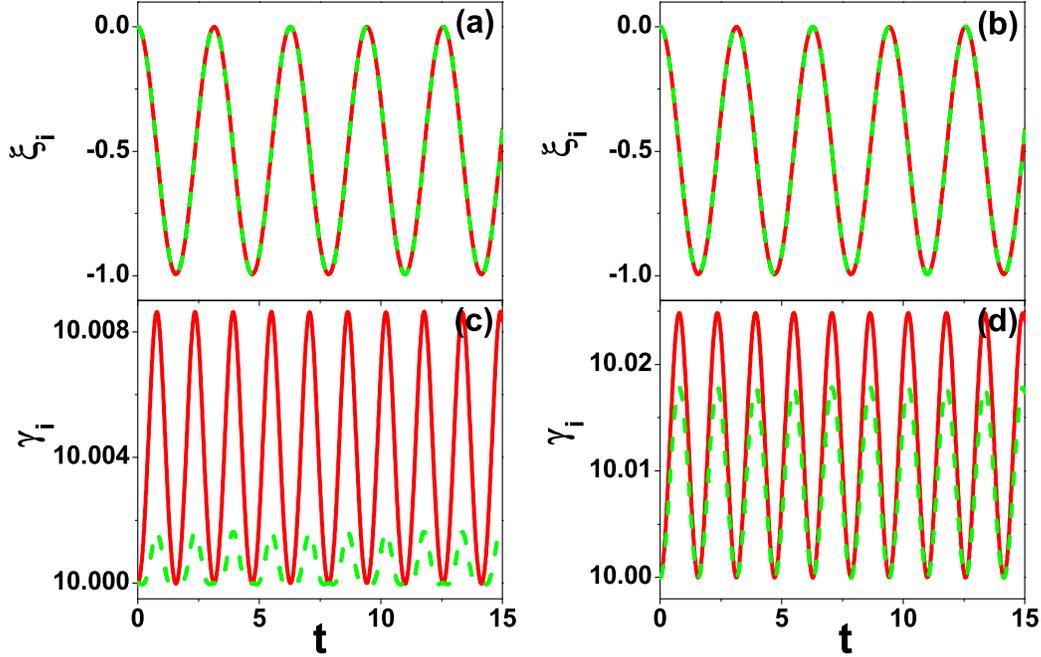}
\caption{(Color online) The stabilization of BOs for different
relative separations $\tau$ under the $\cos(\omega t)$ modulation.
$\tau=0.1$ for (a, c) and $\tau=0.5$ (b, d). The same other
parameters and notation as Fig. \ref{fig:5}.} \label{fig:6}
\end{figure*}

However, in many situations, such finite enhancement to the dephasing
time of BOs is not enough, and one is desired to preserve the BOs
forever. This actually can also be achieved by modulating the
interactions by a suitable time-dependent field. Fig. \ref{fig:6}
shows that the BOs are entirely stabilized by modulating
interactions harmonically in time with the same frequency as the one
of the BOs. Such stabilization is independent of the magnitudes of
the nonlinearities, so the behaviors for different $\tau$ under the
modulation are same, as shown in Fig. \ref{fig:6}(a,b).

\section{Conclusions}
\label{4} In summary, we have studied analytically and numerically
the dynamical behaviors of the BOs for two-component BECs trapped in
combined potentials consisting of linear potentials and
spin-dependent optical lattices. We found that the damped BOs can be
stabilized when all the atom-atom interactions are modulated
synchronously and harmonically in time with Bloch frequency.
Moreover, if the intercomponent and the intracomponent interactions
have inverse signs, it has been shown that the dephasing time of BOs
can be much enhanced by decreasing the relative separation of the
two potentials felt by the two components. Our results provide a
valuable guidance for achieving long-lived BOs in the two-component
BEC system by the Feshbach resonances and manipulating the
configuration of the optical lattices.

\section*{Acknowledgements}
 This work was supported by the Fundamental Research Fund
for Physics and Mathematics of Lanzhou University (Grant
LZULL200806). JHA thanks the support by the Fundamental Research Funds for the Central Universities under Grant
No. lzujbky-2010-72 and the Gansu Provincial NSF of China under Grant No.
0803RJZA095.

\end{document}